\documentclass[aps,prd,preprint,showpacs,preprintnumbers,superscriptaddress,amsmath]{revtex4}

\def\slashed#1{\slash\hskip-6pt #1}

\begin{document}

\title{New-Physics scenarios in $\Delta S = 2$ decays of the $B_c$ meson}

\author{Svjetlana Fajfer}
\email[E-mail:]{svjetlana.fajfer@ijs.si}
\affiliation{J. Stefan Institute, Jamova 39, P. O. Box 3000, 1001 Ljubljana, Slovenia}
\affiliation{Department of Physics, University of Ljubljana, Jadranska 19, 1000 Ljubljana, Slovenia}

\author{Jernej Kamenik}
\email[E-mail:]{jernej.kamenik@ijs.si}
\affiliation{J. Stefan Institute, Jamova 39, P. O. Box 3000, 1001 Ljubljana, Slovenia}

\author{Paul Singer}
\email[E-mail:]{phr26ps@physics.technion.ac.il}
\affiliation{Department of Physics, Technion - Israel Institute of Technology, Haifa 32000, Israel}

\date{\today}

\begin{abstract}
The $\Delta S = 2$ transition $b \rightarrow s s \bar d$ is extremely small in the standard model, proceeding only via box diagrams. Such decays are thus an appropriate ground for searching new physics. We present a study of several two-body and three-body decays of the $B_c$ meson belonging to this class. Our calculation covers the minimal supersymmetric model with and without $\mathcal R$-parity and a two-Higgs-doublet model. The current limits on the parameters of these models allow for rather large branching ratios for several two and three-body hadronic decay modes of $B_c$, up to the $10^{-4}$ range in the minimal supersymmetric model with $\mathcal R$-parity violation.
\end{abstract}


\maketitle

\section{Introduction}

Rare decays of B-mesons are considered to be one of the
promising areas for the discovery of new physics beyond the 
Standard Model(SM)~\cite{Grossman:2003qi,Isidori:2004rd, Buras:2004sc}. 
This is based on the expectation 
that virtual new particles will affect these decays, in 
particular in processes induced by flavor changing neutral 
currents; such processes are suppressed in the SM since they 
proceed via loop diagrams. This venue, typified by transitions 
like $b\rightarrow s(d) \gamma$, $b \rightarrow s(d) l \bar l $ 
has been investigated intensively in recent years~\cite{Grossman:2003qi,
Isidori:2004rd, Buras:2004sc}.
%, both theoretically and experimentally.

\par

A related, though alternative approach is the search for rare 
$b$ decays which have extremely small rates in the SM, and 
their mere detection would be a sign for new physics. Several 
years ago, Huitu, Lu, Singer and Zhang have suggested 
\cite{Huitu:1998vn, Huitu:1998pa} the study of $b \rightarrow s 
s \bar d$ and $b \rightarrow d d \bar s$ as prototypes of the 
alternative method. This proposal is based~\cite{Huitu:1998vn} 
on the fact that these transitions are indeed exceedingly small 
in the SM, where they occur via box diagrams with up-type quarks 
and $W$'s along the box. The matrix elements of these transitions 
are approximately proportional in SM to $\lambda^7$ and 
$\lambda^8$ ($\lambda$ is the sine of the Cabibbo angle), 
resulting in branching ratios of approximately $10^{-11}$ 
and $10^{-13}$ respectively, probably too small for detection 
even at LHC. Further discussions on these ($\Delta S=2$) and 
($\Delta S= -1$) transitions are given in Refs. 
\cite{Grossman:1999av, Fajfer:2001ht, Chun:2003rg, Wu:2003kp}.

\par

Huitu et al. have then investigated~\cite{Buras:2004sc,Huitu:1998vn} 
the $b \rightarrow s s \bar d$ transition in several models of 
physics beyond SM, namely the minimal supersymmetric standard 
model (MSSM), the minimal supersymmetric standard model with 
$\mathcal R$-parity violating couplings (RPV) and Two Higgs Doublet models. 
Within a certain range of allowed parameters, MSSM predicts 
\cite{Huitu:1998vn} a branching ratio as high as $10^{-7}$ for 
$b \rightarrow s s \bar d$. On the other hand, Higgs models may 
lead~\cite{Huitu:1998pa} to a branching ratio at the $10^{-8}$ 
level. Most interestingly, in RPV the process can proceed as a 
tree diagram and the then existing limits~\cite{Huitu:1998vn} 
on $\lambda'_{ijk}$  RPV superpotential couplings did not 
constrain at all the $b \rightarrow s s \bar d$ transition.

\par

In Ref.~\cite{Huitu:1998vn} the hadronic decay 
$B^- \rightarrow K^- K^- \pi^+$, proceeding either directly or 
through a $\bar K^{*0}$, has been selected as a convenient 
signal for the $b \rightarrow s s \bar d$ transition. 
A rough estimate~\cite{Huitu:1998vn} has shown that the 
semi-inclusive decays 
$B^- \rightarrow K^- K^- + \text{(nonstrange)}$  may account for 
about $1/4$ of the $b \rightarrow s s \bar d$ transitions. 
The OPAL collaboration has undertaken  the search for this 
decay establishing~\cite{Abbiendi:1999st} the first upper limit 
for it, $\mathcal{BR}(B^- \rightarrow K^- K^- \pi^+) < $ 
$1.3 \times 10^{-4}$. This result then imposes more severe 
limits on the combination of $\lambda'_{ijk}$ couplings which 
appears in the amplitude for $b \rightarrow s s \bar d$ in RPV, 
than previously obtained from other processes. Very recently, 
the BELLE experiment has reduced~\cite{Abe:2002av,Garmash:2003er} 
the above limit to 
$\mathcal{BR}(B^- \rightarrow K^- K^- \pi^+) <$ $2.4 \times 10^{-6}$. In an experiment planned~\cite{Damet:2000ct} by ATLAS at LHC, a two orders of magnitude improvement is expected.

\par

In order to use these results for restricting the size of the 
$b \rightarrow s s \bar d$ transition one needs also an estimate 
for the long distance (LD) contributions to such ($\Delta S=2$) 
processes. A calculation~\cite{Fajfer:2000ax} of such 
contributions provided by $D D$ and by $D \pi$ intermediate states 
 as well as those induced by virtual $D$ and $\pi$ mesons leads 
 to a branching ratio $\mathcal{BR}^{LD} 
 (B^- \rightarrow K^- K^- \pi^+ ) = 6 \times 10^{-12}$, 
 only slightly larger than the short-distance result of SM for this transition. 
 Thus, this decay and similar ones are indeed suitable for the 
 search of new physics, the LD contribution not overshadowing the 
new physics,if it leads to rates of the order of $10^{-10}$  or larger. 
 A survey~\cite{Fajfer:2000ny} of various 
 possible two-body ($\Delta S=2$) decays of $B$ mesons to 
 $V V$, $V P$, $P P$ states has singled out 
 $B^- \rightarrow K^{*-} \bar K^0$ and 
 $B^- \rightarrow K^- \bar K^0$ as the most likely 
 ones for the detection of the presence of RPV transitions 
 at the $10^{-7}$ level.

\par

In the present paper we study the ($\Delta S=2$) transitions of 
the $B_c$ meson. This double-heavy meson, which has been firstly 
observed~\cite{Abe:1998wi, Abe:1998fb} by CDF Collaboration a 
few years ago, can decay weakly by b-quark decay with the 
$c$-quark as spectator, by $c$-quark decay with the $b$-quark 
as spectator or via an annihilation process. There is already a 
fairly large number of theoretical papers describing the various 
leptonic, semileptonic and hadronic exclusive channels to 
which $B_c$ may decay in the Standard Model. 
These are well summarized in a recent review by Kiselev 
\cite{Kiselev:2003mp}.

\par

At the forthcoming LHC accelerator one expects about 
$5 \times 10^{10}$ $B_c$ events/year, at a luminosity of 
$10^{34}\mathrm{~cm}^{-2} s^{-1}$~\cite{Gouz:2002kk}. 
Hence, the search for rare decays of $B_c$ exhibiting possible 
features of physics beyond SM will also become possible. 
The effect of such physics on radiative decays of $B_c$, as 
caused for example by $c$-quark decay via the 
$c \rightarrow u + \gamma$ transition has already been 
investigated~\cite{Fajfer:1999dq, Aliev:1999tg}. 
Here, we address the effects of new physics on rare $b$-decays caused by 
 the ($\Delta S=2$) transitions, which we already mentioned to be very rare in SM~\cite{Huitu:1998vn, Huitu:1998pa, Grossman:1999av}. Specifically, we calculate two-body decay modes $B_c^- \rightarrow D^{*-}_s \bar K^{*0}$, $D^{*-}_s \bar K^0, D_s^- \bar K^{*0}$ and $D_s^- \bar K^0$ as well as three body modes  $B_c^- \rightarrow D_s^- K^- \pi^+$, $D^{*-}_s K^- \pi^+$, $D_s^-  D^{*-}_s D^+$, $D_s^-  D_s^- D^{*+}$, $D_s^-  D_s^- D^+$, $D^0 \bar K^0 K^-$ and $D^{*0} \bar K^0 K^-$. We expect these modes to be most likely candidates for the experimental observation. Our calculation shows that while the SM prediction for the branching ratio for these exclusive channels is $10^{-13}$ or less, some of the ($\Delta S=2$) decays can have branching ratios as high as $10^{-6} - 10^{-4}$ in MSSM with RPV.

\par

In the second section we describe the framework used in our 
calculation and the beyond SM models which we consider. 
In Section 3 we give the explicit expressions for the 
transition matrix elements. In Section 4 we discuss the 
results and the prospects for checking the "beyond SM physics" 
in $B_c$ decays.

\section{Framework}

The Standard Model analysis describing the $b \rightarrow s s \bar d$ transition 
leads to the 
effective Hamiltonian:~\cite{Huitu:1998vn}
\begin{equation}
  \mathcal{H}_{SM} = \widetilde C_{SM} \left( \bar s_L \gamma^{\mu} d_L \right)
  \left( \bar s_L \gamma_{\mu} b_L \right),
\end{equation}
where we have denoted
\begin{eqnarray}
	\widetilde C_{SM} &=& \frac{1}{2}\Bigg\{\frac{G_F^2}{2\pi^2} m_W^2 V_{tb} V^*_{ts} \Bigg[ V_{td} V_{ts}^* f\left( \frac{m_W^2}{m_t^2} \right) \nonumber\\*
	&& + V_{cd} V_{cs}^* \frac{m_c^2}{m_W^2} g\left( \frac{m_W^2}{m_t^2},\frac{m_c^2}{m_W^2} \right) \Bigg]\Bigg\},
\end{eqnarray}
with functions $f(x)$ and $g(x,y)$ given in~\cite{Huitu:1998vn}. 
Taking numerical values from PDG~\cite{Hagiwara:2002fs} and neglecting 
the CKM phases, one estimates 
$|C_{SM}|\simeq 4\times10^{-12} \mathrm{~GeV}^{-2}$.

\par

Next we turn to several models beyond SM; the minimal supersymmetric SM, the minimal supersymmetric SM with $\mathcal R$-parity violation and a two-Higgs-doublet model. The largest contribution found in the 
minimal supersymmetric extension of the standard model 
\cite{Huitu:1998vn} gives for the corresponding factor
\begin{equation}
  \widetilde C_{MSSM} = -\frac{\alpha_S^2 \delta^{*}_{12}
  \delta_{23}}{216m_{\widetilde
  d}^2}\left[ 24x f_6 (x) + 66 \widetilde f_6 (x) \right],
\end{equation}
with $\alpha_S(m_W)=0.12$, $x=m_{\widetilde g}^2/m_{\widetilde d}^2$, and the functions $f_6(x)$ and $\widetilde f_6 (x)$ given in~\cite{Gabbiani:1996hi}. The couplings $\delta^d_{ij}$ parametrize the mixing between the down-type left-handed squarks. The value of $\delta_{12}$ is determined from the $K^0-\bar K^0$ mixing~\cite{Gabbiani:1996hi} and is currently bounded by $m_{K_L}-m_{K_S}=3.49\times10^{-15}\mathrm{~GeV}$~\cite{Hagiwara:2002fs}. We present our calculation choosing $x=1$ and $8$; using results from~\cite{Gabbiani:1996hi} we estimate the absolute value of $\delta_{12}$ to be below $3(9)\times 10^{-2}$ at average squark mass $m_{\tilde d}=350(500)\mathrm{~GeV}$ and $x=1(8)$. The strongest bounds on $\delta_{23}$ come from the radiative $b\rightarrow s\gamma$ decay ~\cite{Gabbiani:1996hi, Ciuchini:2003sq,Ciuchini:2003rg}. These studies give at $x=1$ and for $m_{\tilde d}=350\mathrm{~GeV}$ the stronger bound on $|\delta_{23}|\lesssim 0.4$ which results in $|\widetilde C_{MSSM}| \lesssim 5 \times 10^{-12} \mathrm{~GeV}^{-2}$. Taking the same values as in~\cite{Fajfer:2000ny}, $x=8$, $m_{\tilde d} = 500\mathrm{~GeV}$, leads to $|\widetilde C_{MSSM}| \lesssim 3 \times 10^{-10} \mathrm{~GeV}^{-2}$. These updated values for $\widetilde C_{MSSM}$ are somewhat smaller than those used in~\cite{Huitu:1998vn,Fajfer:2000ny}.

%The next to leading order calculation at $x=1$ and $m_{\tilde d}=350\mathrm{~GeV}$~\cite{Ciuchini:2003rg} bounds the value below the order of $\sim 0.4$. At $x=8$, beyond leading order corrections are not yet available, and the leading order calculation from~\cite{Gabbiani:1996hi} at $m_{\tilde d}=500\mathrm{~GeV}$ gives a value of $\sim 40$. We thus estimate at the scale of $b$ quark mass for the coupling at $x=1$ and $x=8$ to be $|\widetilde C_{MSSM}| \lesssim 5 \times 10^{-12} \mathrm{~GeV}^{-2}$ and $|\widetilde C_{MSSM}| \lesssim 1 \times 10^{-10} \mathrm{~GeV}^{-2}$ respectively.

\par

When including $R$-parity violating interactions in the MSSM, the 
part of the superpotential which is relevant here is 
$W=\lambda'_{ijk} L_i Q_j d_k$, where $i$, $j$ and $k$ 
are indices for the families and the $L$, $Q$ and $d$ are 
superfields for the lepton doublet, the quark doublet and the 
down-type quark singlet, respectively. Using the notation of 
\cite{Choudhury:1996ia} and~\cite{Huitu:1998vn}, the tree 
level Hamiltonian has the form
\begin{eqnarray}
	\mathcal H_R &=& - \sum_{n=1}^{3} \frac{f_{QCD}}{m_{\widetilde \nu_n}^2} \big[
  \lambda'_{n32} \lambda_{n21}'^* (\bar s_R b_L)( \bar s_L d_R) \nonumber\\* 
	&& + \lambda'_{n12} \lambda_{n23}'^* (\bar s_R d_L)( \bar s_L b_R) \big].
\label{H_R}
\end{eqnarray}
The QCD corrections were found to be important for this 
transition~\cite{Bagger:1997gg}. For our purpose it suffices 
to follow~\cite{Huitu:1998vn} retaining the leading order QCD 
result $f_{QCD}\simeq 2$, for 
$m_{\widetilde \nu}=100\mathrm{~GeV}$.

\par

The most recent upper bound on the specific combination of 
couplings entering (\ref{H_R}) can be obtained from BELLE's 
search for the $B^+\rightarrow K^+ K^+ \pi^-$ decay 
\cite{Abe:2002av,Garmash:2003er}. The result based on the explicit calculation of the amplitude $B^-\rightarrow K^- K^- \pi^+$~\cite{Fajfer:2001te} 
\begin{eqnarray}
&& A^{RPV}_{B^- \rightarrow K^- K^- \pi^+} = \sum_{n=1}^{3} \frac{f_{QCD}}{m_{\widetilde \nu_n}^2} \left[ \lambda'_{n32} \lambda_{n21}'^* + \lambda'_{n12} \lambda_{n23}'^*\right] \nonumber\\
&& \quad  \times F_0^{K\pi}(q^2) F_0^{BK}(q^2)\frac{(m_B^2-m_K^2)(m_K^2-m_{\pi}^2)}{(m_s-m_d)(m_b-m_s)}
\end{eqnarray} 
and experimental upper bound from~\cite{Garmash:2003er} which corresponds to branching ratio $\mathcal{BR}(B^-\rightarrow K^- K^- \pi^+) < 2.4 \times 10^{-6}$ is $\sqrt{\sum_{n=1}^{3} \left|\lambda'_{n32} \lambda_{n21}'^*\right|^2 + \left|\lambda'_{n21} \lambda_{n32}'^* \right|^2 }<  $ $3.7 \times 10^{-4}$. Note that this is a somewhat larger value than previous estimates ~\cite{Fajfer:2001te,Fajfer:2000ny} based on the assumption of $\Gamma(B^-\rightarrow K^- K^- \pi^+)\simeq 1/4 \Gamma(b\rightarrow s s \bar d)$~\cite{Huitu:1998vn}. 
%This is because explicit calculation of the branching ratio for the decay $B^-\rightarrow K^- K^- \pi^+$ from~\cite{Fajfer:2001te} significantly alters the ratio $\Gamma(B^-\rightarrow K^- K^- \pi^+)/\Gamma(b\rightarrow s s \bar d)$ which we now assume to be of the order of $1/10$.

\par

The decay $b\rightarrow s s \bar d$ has been investigated using two Higgs doublet models as well~\cite{Huitu:1998pa}. These authors found that the charged Higgs box contribution in MSSM is negligible. In two models with charged Higgs branching ratios are very small~\cite{Huitu:1998pa} for reasonable values of $\cot \beta$. Recently it has been also verified~\cite{Wu:2003kp} that the chargino contribution in MSSM to this process is indeed smaller by an order of magnitude than contributions calculated in~\cite{Huitu:1998vn}. On the other hand, THDM involving several neutral Higgs bosons~\cite{Cheng:1987rs} could in principle give larger inclusive decay rates for $b \to s s \bar d$. The part of the effective Hamiltonian relevant in our case is the tree diagram exchanging the neutral Higgs bosons $h$ (scalar) and $A$ (pseudoscalar)
\begin{equation}
	\mathcal H_{TH} = \frac{i}{2} \xi_{sb} \xi_{sd} \left[ \frac{1}{m_h^2} (\bar s d) (\bar s b) - \frac{1}{m_A^2} (\bar s \gamma_5 d) (\bar s \gamma_5 b) \right],
\end{equation}
with the coupling $\xi_{ij}$ defined in~\cite{Cheng:1987rs} as a Yukawa coupling of the flavor changing neutral current transitions $d_i \leftrightarrow d_j$. In our estimation we use the bound $|\xi_{sb}\xi_{sd}|/m_H^2>10^{-10} \mathrm{~GeV^{-2}}$, $H=h,A$ which was obtained in~\cite{Huitu:1998pa} by using the $\Delta m_K$ limit on $\xi_{bd}/m_H$ and assuming $|\xi_{sb}/m_H|>10^{-3}$.

\par

In order to estimate the amplitudes for the various $\Delta S = 2$ exclusive decays of $B_c$, 
one has to calculate the matrix elements of the operators 
appearing in the effective Hamiltonian.  We rely on the factorization 
approximation~\cite{Ali:1998eb, Wirbel:1985ji, Bauer:1987bm} 
which requires the knowledge of matrix elements of the current 
operators or density operators. More sophisticated methods such as perturbative QCD calculation have been employed in the $B^0 \to \Phi \Phi$ case~\cite{Bar-Shalom:2002sv}. However, in the case of nonleptonic weak decays of $B_c$ mesons such approaches have not yet been developed. We use the standard form factor representation of hadronic 
current matrix elements~\cite{Ali:1998eb, Ebert:2004ka, 
Ebert:2003cn, Kiselev:2002vz, Bauer:1987bm, Wirbel:1985ji}. 
For the matrix element between two pseudoscalar mesons we employ
\begin{eqnarray}
	&&\langle P_2 (p_2) | \bar q_j \gamma^{\mu} q_i | P_1 (p_1) \rangle\nonumber\\*
	&&\quad= f_+ (q^2) \left(p_1+p_2\right)^{\mu} + f_-(q^2) q^{\mu} \nonumber\\*
	&&\quad= F_1(q^2) \left( (p_1+p_2)^{\mu} - \frac{m_{P_1}^2-m_{P_2}^2}{q^2}q^{\mu} \right) \nonumber\\*
	&&\quad\quad+ F_0(q^2) \frac{m_{P_1}^2-m_{P_2}^2}{q^2}q^{\mu},
\end{eqnarray}
where $q^{\mu} = (p_1-p_2)^{\mu}$ and the two different parameterizations are connected via $F_0(q^2) = f_+(q^2) + f_-(q^2) q^2/(m_{P_1}-m_{P_2})$ and $F_1(q^2) = f_+(q^2)$. The matrix element between a pseudoscalar and vector meson we decompose as customary
\begin{eqnarray}
	&&\langle V (\epsilon_V,p_2) | \bar q_j \gamma^{\mu}(1-\gamma^5) q_i | P (p_1) \rangle\nonumber\\* 
	&&\quad= \frac{2 V(q^2)}{m_P + m_V} \epsilon^{\mu\nu\alpha\beta} \epsilon_{V\nu}^* p_{1\alpha} p_{2\beta} - i \epsilon_V^* \cdot q \frac{2m_V}{q^2} q^{\mu} A_0(q^2) \nonumber\\*
	&&\quad\quad- i(m_P + m_V) \left[\epsilon_V^{*\mu} - \frac{\epsilon^{*} \cdot q}{q^2} q^{\mu}\right] A_1(q^2) \nonumber\\*
	&&\quad\quad+ i \frac{\epsilon_V^{*}\cdot q}{(m_P + m_V)}\left[(p_1+p_2)^{\mu} - \frac{m_P^2-m_V^2}{q^2} q^{\mu}\right] A_2(q^2)  \nonumber\\*
	&&\quad= - F_V(q^2) \epsilon^{\mu\nu\alpha\beta} \epsilon^*_{V\nu} p_{1\alpha} p_{2\beta} + i F_0^A(q^2) \epsilon^{*\mu} \nonumber\\*
	&&\quad\quad  + i F_+^A(q^2) \epsilon^*_V \cdot p_1 (p_1+p_2)^{\mu} + i F_-^A \epsilon^*_V \cdot p_1 q^{\mu}, 
\end{eqnarray}
and again the two different parameterizations are connected by $V(q^2) = F_V(q^2) (m_P+m_V)$, $A_0(q^2) = (1/2m_V) (-F^A_0(q^2) + F^A_+(q^2) (m_P^2-m_V^2) + F^A_-(q^2) q^2)$, $A_1(q^2) = F^A_0(q^2)/(m_P+m_V)$ and $A_2(q^2) = F^A_+(q^2) (m_P+m_V)$. We also use the following decay constants
\begin{subequations}
\begin{eqnarray}
\langle P (p) | \bar q_j \gamma^{\mu}\gamma^{5} q_i | 0 \rangle &=& i f_P p^{\mu},\\
\langle S (p) | \bar q_j \gamma^{\mu} q_i | 0 \rangle &=& f_S p^{\mu},\\
\langle V (\epsilon_V,p) | \bar q_j \gamma_{\mu} q_i | 0 \rangle &=& \epsilon_{\mu}^* g_V(p^2),
\end{eqnarray}
\end{subequations}
Finally, for the calculations of the density operators we use relations
\begin{subequations}
\begin{eqnarray}
\partial^{\alpha} (\bar s \gamma_{\alpha} b) &=& i (m_b-m_s) \bar s b \\
\partial^{\alpha} (\bar s \gamma_{\alpha}\gamma_5 b) &=& i (m_b+m_s) \bar s\gamma_5 b.
\end{eqnarray}
\end{subequations}

\par

In our calculations we need the $B_c\rightarrow D_s^{(*)}$ transition form factors $f_{\pm}$, $F_V$, $F_0^{A}$  and $F_{\pm}^{A}$. Since heavy quark effective theory (HQET) is not directly applicable to the decays of the $B_c$ meson, we assume pole dominance for these form factors~\cite{Wirbel:1985ji,Kiselev:2002vz}:
\begin{eqnarray}
F(q^2) &=& \frac{F(0)}{(1 - q^2/m_{\mathrm{pole}}^2)}
\end{eqnarray}
and take numerical values for $F(0)$ and $m_{\mathrm{pole}}$ from from QCD sum rules calculations~\cite{Kiselev:2002vz} (see Tables \ref{F0_table} and \ref{M_table}). 
\begin{table}%[H] add [H] placement to break table across pages
\caption{\label{F0_table} Numerical values of $B_c\rightarrow D_s$ transition form factors at $q^2=0$ taken from~\cite{Kiselev:2002vz}.}
\begin{ruledtabular}
\begin{tabular}{cccccc}
	$f_+$ & $f_-$ & $F_V\mathrm{~(GeV)}^{-1}$ & $F_0^A\mathrm{~(GeV)}$ & $F_+^A\mathrm{~(GeV)}^{-1}$ & $F_-^A\mathrm{~(GeV)}^{-1}$\\\hline
	$0.45$ & $-0.43$ & $0.24$ & $4.7$ & $-0.077$ & $0.13$\\
\end{tabular}
\end{ruledtabular}
\end{table}
\begin{table}%[H] add [H] placement to break table across pages
 \caption{\label{M_table} Pole masses used in $B_c\rightarrow D_s$ transition form factors taken from~\cite{Kiselev:2002vz}.}
 \begin{ruledtabular}
 \begin{tabular}{cccccc}
	$f_+ (\mathrm{GeV})$ & $f_- (\mathrm{GeV})$ & $F_V (\mathrm{GeV})$ & $F_0^A (\mathrm{GeV})$ & $F_+^A (\mathrm{GeV})$ & $F_-^A (\mathrm{GeV})$\\\hline
	$5.0$ & $5.0$ & $6.2$ & $\infty$ & $6.2$ & $6.2$\\
 \end{tabular}
 \end{ruledtabular}
 \end{table}
For the $F_1^{K\pi}$ and $F_0^{K\pi}$, which are needed in the three body decay modes involving $K^-$ and $\pi^+$ mesons we use results from~\cite{Fajfer:1999hh}
\begin{eqnarray}
	F_1^{K\pi}(q^2) &=&  \frac{2 g_{V K(892)}  
	g_{K^*}}{q^2-m_{K(892)}^2+ i \sqrt{q^2} \Gamma_{K(892)}}\\
	F_0^{K\pi}(q^2) &=&  \frac{2 g_{V K(892)}  
	g_{K^*}(1-q^2/m_{K(892)}^2)}{q^2-m_{K(892)}^2+ 
	i \sqrt{q^2} \Gamma_{K(892)}} \nonumber\\* 
	&& \hskip-60pt +\frac{q^2}{(m_K^2-m_{\pi}^2)} 
	 \frac{f_{K(1430)} g_{S K(1430)}}{q^2-m_{K(1430)}^2 + i \sqrt{q^2} \Gamma_{K(1430)}}
\end{eqnarray}
In our numerical calculations we use the following values: 
$f_K = 0.162\mathrm{~GeV}$, $f_{K^0} (1430) \simeq 0.05\mathrm{~GeV}$, $g_{K^*}= 0.196\mathrm{~GeV}^2$, $g_{V K(892)}=4.59  $ and $g_{S K(1430)} = 
3.67 \pm 0.3\mathrm{~GeV}$ taken from~\cite{Fajfer:1999hh}. 

\par

In the three body decay modes involving pairs of $D$ and 
$D_s$ mesons, we also need the form factors for the 
$D_s^{(*)} \rightarrow D^{(*)}$ transitions. These are not available in the literature and we calculate them by utilizing HQET, including the light scalar meson interactions with heavy mesons as it has been done recently~\cite{Bardeen:2003kt}, and presuming the main contributions from exchange of light meson resonances 
($K^0$, $K^*(892)$ and $K^{*0}(1430)$). 
Interactions of heavy mesons are described by the HQET 
chiral Lagrangian 
\cite{Wise:1992hn, Burdman:1992gh,Casalbuoni:1997pg,Bardeen:2003kt}
\begin{eqnarray}
	&\mathcal L_{S} = i \mathrm{Tr}\left[ H v_{\mu} (\partial + \mathcal V)^{\mu} \overline H \right] + i g_A \mathrm{Tr}\left[ H \gamma_{\mu} \gamma_{5} \mathcal A^{\mu} \overline H \right] \nonumber\\*
	&+ i \tilde \beta \mathrm{Tr}\left[ H v_{\mu} (\mathcal V^{\mu} - \rho^{\mu} ) \overline H \right] - \frac{g_{\pi}}{4} \mathrm{Tr} \left[ H \tilde \sigma \overline H \right] + \ldots
\end{eqnarray} 
We use standard heavy meson field notation 
$H = 1/2 \left(\slashed v + 1\right) (P^*_{\mu}\gamma^{\mu}-P \gamma_5)$, 
where $P$ and $P^*_{\mu}$ create heavy pseudoscalar and vector 
mesons respectively, and exponential representation of light 
pseudoscalar meson fields $\xi = \exp(i \mathcal M / f)$, 
where $\mathcal M$ is the light pseudoscalar meson  
chiral $SU(3)$ nonet matrix, so that 
$\mathcal V_{\mu} = 1/2 (\xi^{\dagger} \partial_{\mu} \xi + 
\xi \partial_{\mu} \xi^{\dagger})$, 
$\mathcal A_{\mu} = 1/2 (\xi^{\dagger} \partial_{\mu} \xi - 
\xi \partial_{\mu} \xi^{\dagger})$. 
Light vector fields are represented as 
$\rho_{\mu} = i g_V \hat\rho_{\mu} /\sqrt 2 $, 
where $\hat\rho_{\mu}$ is the light vector meson chiral $SU(3)$ 
octet matrix. Finally light scalar mesons are introduced 
through the $\tilde \sigma=\sqrt{2/3} \sigma$ field, where 
$\sigma$ is the light scalar meson chiral $SU(3)$ nonet matrix. 
The ellipses indicate further terms involving only light meson 
fields, chiral and $1/m_H$ corrections. 
The light vector meson coupling ($\tilde \beta$) was found to be 
consistent with zero~\cite{Bajc:1997ey}. 
Consequently we neglect the contribution of the vector $K^*(892)$ 
resonance in our calculations. 
In the HQET approximation in the leading order in heavy quark 
mass and chiral perturbation $f_+^{D_sD}$, 
$F_V^{D_s^*D}$, $F_V^{D_sD^*}$, 
$F_0^{A D_s^*D}$, $F_0^{A D_sD^*}$, 
$F_+^{A D_s^*D}$ and $F_+^{A D_sD^*}$ 
are all found to vanish, so the only contributions come to the 
$f_-^{D_sD}$, $F_-^{A D_s^*D}$ and 
$F_-^{A D_sD^*}$ form factors
\begin{eqnarray}
	f_-^{D_s D}(q^2) &=&  
	\frac{(g_{\pi}/4) f_{K(1430)} \sqrt{m_{D_s}m_D} }{q^2-m_{K(1430)}^2 + i \sqrt{q^2} \Gamma_{K(1430)}}\\
	F_-^{A D_s^*D}(q^2) &=& F_-^{A D_sD^*}(q^2) = 2 g_A \frac{\sqrt{m_{D_s} m_D}}{q^2 - m_K^2}.\nonumber\\
\end{eqnarray}
In the numerical calculation we use $g_\pi =\simeq 3.73$ 
and $g_A \simeq 0.59$~\cite{Bardeen:2003kt}. The rest of parameters are 
taken from PDG~\cite{Hagiwara:2002fs}.

\section{Amplitudes}

\subsection{Two-body decays}

First we investigate two-body decays of $B_c^-$ mesons which a priori are the
promissing candidates for the experimental studies.

\subsubsection{$B_c^- \rightarrow D_s^{*-} \bar K^{*0}$ decay}

For the analysis of pseudoscalar meson decay into two vector 
mesons it is convenient to use partial helicity formalism 
(see, e.g.,~\cite{ElHassanElAaoud:1999nx}).
The amplitude can be decomposed in the product 
of the effective Wilson coefficient and the matrix element 
of the operator. Within SM we have 
 \begin{equation}
	\mathcal M(B_c^- \rightarrow D_s^{*-} \bar K^{*0}) = \frac{1}{4} C_{SM} \langle D_s^{*-} \bar K^{*0} | \mathcal O | B_c^- \rangle,
\end{equation}
where $	\mathcal O = (\bar s \gamma^{\mu} (1-\gamma_5) d) (\bar s \gamma^{\mu} (1-\gamma_5) b) $. Using factorization one finds $\langle D_s^{*-} \bar K^{*0} | \mathcal O | B_c^- \rangle$ $=$ $- \langle D_s^{*-} | (\bar s \gamma^{\mu} b) | B_c^- \rangle$ $\times \langle \bar K^{*0} | (\bar s \gamma^{\mu} \gamma_5 d) | 0 \rangle$ which by using helicity formalism leads to the following expressions for amplitudes:
\begin{eqnarray}
	\mathcal M_{00} &=& \frac{1}{4}C_{SM} g_{K^*} (m_{B_c} + m_{D_s^{*}}) \nonumber\\*
	&&\times [ \alpha A_1 (m_{K^*}^2) - \beta A_2 (m_{K^*}^2) ], \\
	\mathcal M_{\pm} &=& \frac{1}{4}C_{SM} g_{K^*} (m_{B_c} + m_{D_s^{*}}) \nonumber\\*
	&&\times [ \alpha A_1 (m_{K^*}^2) \mp \gamma V (m_{K^*}^2) ],
\end{eqnarray} 
where $\alpha = (1-r^2-t^2)/2rt$, $\beta=k^2/(2 r t (1+r^2))$, $\gamma=k/(1+r^2)$ with $r=m_{D_s^{*}}/m_{K^*}$, $t=m_{K^*}/m_{B_c}$ and $k=\sqrt{1+r^4+t^4-2r^2-2t^2-2 r^2 t^2}$. The decay with is then
\begin{eqnarray}
	&&\Gamma(B_c^- \rightarrow D_s^{*-} \bar K^{*0}) \nonumber\\*
	&& \quad= \frac{|\vec p|}{8 \pi m_{B_c}^2}  \left[|\mathcal M_{00}|^2+|\mathcal M_{++}|^2+|\mathcal M_{--}|^2\right].
\label{BDsKs}
\end{eqnarray} 
Within the MSSM the result in (\ref{BDsKs}) can be reused by replacing $C_{SM}$ with $C_{MSSM}$. On the other hand RPV and THDM models cannot be tested in this mode when factorization approach is used due to the vanishing $\langle \bar K^{*0}(p_K) | (\bar s \gamma^{\mu} (1-\gamma_5) d) | 0 \rangle$ matrix element.

\subsubsection{$B_c^- \rightarrow D_s^{*-} \bar K^0$ decay}

In SM and MSSM factorization gives for the $\mathcal O$ matrix element $\langle D_s^{*-} \bar K^0 | \mathcal O | B_c^- \rangle = \langle \bar K^0 | (\bar s \gamma^{\mu} (1-\gamma_5) d) | 0 \rangle \langle D_s^{*-}  | (\bar s \gamma^{\mu} (1-\gamma_5) b) | B_c^- \rangle = - 2 i m_{D_s^{*}} f_K A_0(m_K^2) \epsilon^*_D \cdot p_K$. The decay width becomes
\begin{eqnarray}
	&&\Gamma(B_c^- \rightarrow D_s^{*-} \bar K^0) \nonumber\\*
	&&\quad= \frac{|\vec p|}{8 \pi m_{B_c}^2}  \left|\frac{1}{4} C_{SM}\right|^2 \sum_{\mathrm{pol.}} \left|\langle D_s^{*-} \bar K^0 | \mathcal O_{1} | B_c^- \rangle
  \right|^2,
\end{eqnarray} 
where $\sum_{\mathrm{pol.}}$ sums over all polarizations of $D_s^{*-}$ meson. One finds $\sum_{\mathrm{pol.}} |\epsilon^*_D \cdot p_K|^2 = \lambda(m_{B_c}^2,m_K^2,m_{D_s^{*}}^2)$, where $\lambda(a,b,c) = (a^2+b^2+c^2-2(a b + b c + a c))/4c$. RPV can also occur in this decay. The decay amplitude is
\begin{eqnarray}
	&&\mathcal M(B_c^- \rightarrow D_s^{*-} \bar K^0) = -\sum_n \frac{f_{QCD}}{m^2_{\tilde \nu_n}} \nonumber\\*
	&&\quad\times\Bigg[ \frac{\lambda'_{n32} \lambda'^*_{n21}}{4} \langle D_s^{*-} \bar K^0 | \mathcal O_{R_1} | B_c^- \rangle \nonumber\\*
	&&\quad + \frac{\lambda'_{n21} \lambda'^*_{n32}}{4} \langle D_s^{*-} \bar K^0 | \mathcal O_{R_2} | B_c^- \rangle\Bigg],\nonumber\\
\end{eqnarray}
where we have denoted $\mathcal O_{R_1} = \left( \bar s (1-\gamma_5) b \right)\left( \bar s (1+\gamma_5) d \right) $ and $\mathcal O_{R_2} = \left( \bar s (1-\gamma_5) d \right)\left( \bar s (1+\gamma_5) b \right)$. Factorization yields 
\begin{eqnarray}
	&&\langle D_s^{*-} \bar K^0 | \mathcal O_{R_1} | B_c^- \rangle = \langle D_s^{*-} \bar K^0 | \mathcal O_{R_2} | B_c^- \rangle \nonumber\\*
	&&\quad=  \langle \bar K^0 | (\bar s (1-\gamma_5) d) | 0 \rangle \langle D_s^{*-}  | (\bar s (1+\gamma_5) b) | B_c^- \rangle \nonumber\\*
	&&\quad= \frac{m_K^2 f_K}{(m_s+m_d)(m_s+m_b)} \left( 2m_{D_s^{*}} \epsilon^*_D \cdot p_K \right) A_0 (m_K^2),\nonumber\\
\label{DK_R}
\end{eqnarray}
giving for the decay width
\begin{eqnarray}
	&&\Gamma(B_c^- \rightarrow D_s^{*-} \bar K^0)= \frac{|\vec p|}{8 \pi m_{B_c}^2}  \frac{f_{QCD}^2}{m_{\widetilde \nu}^4} \nonumber\\*
	&&\quad\times\left( \sum_n \left|\lambda'_{n32} \lambda_{n21}'^*\right|^2 + \left|\lambda'_{n21} \lambda_{n32}'^* 
  \right|^2 \right) \nonumber\\*
	&&\quad\times\sum_{\mathrm{pol.}} \left|\frac{1}{4} \langle D_s^{*-} \bar K^0 | \mathcal O_{R_1} | B_c^- \rangle
  \right|^2.
\end{eqnarray} 
Similarly, the THDM gives for the amplitude of this decay
\begin{eqnarray}
	&&\mathcal M(B_c^- \rightarrow D_s^{*-} \bar K^0) = -\frac{i}{2} \xi_{sb} \xi_{sd} \nonumber\\*
	&&\quad \times \left[ \frac{1}{m_h^2} \langle D_s^{*-} \bar K^0 | \mathcal O_{T_1} | B_c^- \rangle - \frac{1}{m_A^2} \langle D_s^{*-} \bar K^0 | \mathcal O_{T_2} | B_c^- \rangle\right],\nonumber\\
\end{eqnarray}
where $\mathcal O_{T_1} = (\bar s d) (\bar s b)$ and $\mathcal O_{T_2} = (\bar s \gamma_5 d) (\bar s \gamma_5 b)$. Only the second amplitude turns out to be non-vanishing and gives the same result as (\ref{DK_R}). Because of the specific combination of products of scalar (pseudoscalar) densities, this is the only two body decay which has non-vanishing THDM amplitude within the factorization assumption.

\subsubsection{$B_c^- \rightarrow D_s^- \bar K^{0*}$ decay}

For this decay mode the matrix element of the operator $\mathcal O$ is determined to be $\langle D_s^- \bar K^{*0} | \mathcal O | B_c^- \rangle = \langle \bar K^{*0} | (\bar s \gamma^{\mu} (1-\gamma_5) d) $ $ | 0 \rangle \langle D_s^-  | (\bar s \gamma^{\mu} (1-\gamma_5) b) | B_c^- \rangle = 2 g_K F_1(m_{K^*}^2) \epsilon^*_K \cdot p_D$. On the other hand both RPV and THDM amplitudes for this mode vanish, due to the vanishing of the matrix element of the density operator for $\bar K^{*0}$ state.

\subsubsection{$B_c^- \rightarrow D_s^- \bar K^0$ decay}
 
The matrix element for the operator $\mathcal O$ becomes in this case $\langle D_s^- \bar K^0 | \mathcal O | B_c^- \rangle = \langle \bar K^0 | (\bar s \gamma^{\mu} (1-\gamma_5) d) | 0 \rangle \langle D_s^-  | (\bar s \gamma^{\mu} (1-\gamma_5) b) | B_c^- \rangle = i f_K F_0(m_K^2)(m_{B_c}^2-m_{D_s}^2)$, while for the operators $\mathcal O_{R_1}$ and $O_{R_2}$ we get 
\begin{eqnarray}
	&&\langle D_s^- \bar K^0 | \mathcal O_{R_1} | B_c^- \rangle = -\langle D_s^- \bar K^0 | \mathcal O_{R_2} | B_c^- \rangle \nonumber\\*
	&&\quad= \langle \bar K^0 | (\bar s (1-\gamma_5) d) | 0 \rangle \langle D_s^-  | (\bar s (1+\gamma_5) b) | B_c^- \rangle \nonumber\\*
	&&\quad= -i \frac{m_K^2 f_K}{(m_s+m_d)(m_b-m_s)} F_0 (m_K^2) (m_{B_c}^2-m_D^2).\nonumber\\
\end{eqnarray} 
THDM does not contribute to this mode.

\subsection{Three-body decays}

Next we study three body decay modes of $B_c^-$ which are also possibly good candidates 
for experimental searches.

\subsubsection{$B_c^- \rightarrow D_s^- K^- \pi^+$ decay}

The required matrix element in SM and MSSM is here
\begin{widetext}
\begin{eqnarray}
	&&\langle D_s^- K^- \pi^+| \mathcal O | B_c^- \rangle \nonumber\\*
	&&\quad= \langle K^- \pi^+ | (\bar s \gamma^{\mu} (1-\gamma_5) d) | 0 \rangle \langle D_s^-  | (\bar s \gamma^{\mu} (1-\gamma_5) b) | B_c^- \rangle \nonumber\\*
	&&\quad= f_+^{BD}(s) f_+^{K\pi}(s) (t-u) + f_+^{BD}(s) f_-^{K\pi}(s) ( m_D^2 - m_{B_c}^2 )\nonumber\\*
	&&\quad\quad + f_-^{BD}(s) f_+^{K\pi}(s) (m_{\pi}^2 - m_{K}^2) + f_-^{BD}(s) f_-^{K\pi}(s) (-s),
\end{eqnarray}
\end{widetext}
where $s = m_{23}^2 = q^2 = (p_B-p_D)^2$, 
$u = m_{12}^2 = $ $ (p_B-p_{\pi})^2$ and 
$t=m_{13}^2 = (p_B-p_K)^2$. We obtain the decay 
rate by integrating the transition matrix element over the 
whole Dalitz plot
\begin{eqnarray}
	&&\Gamma(B_c^- \rightarrow D_s^- K^- \pi^+) = \frac{1}{(2\pi)^3} \frac{1}{32 m_{B_c}^3} \nonumber\\*
	&&\quad\times\int dm_{12}^2 \int dm_{23}^2 \left|\frac{1}{4} C_{SM} \langle D_s^- K^- \pi^+ | \mathcal O_{1} | B_c^- \rangle
  \right|^2,\nonumber\\
\end{eqnarray} 
where $(m_{12}^{2})_{max}={(m_{B_c}-m_{\pi})^2}$, $(m_{12}^2)_{min}={(m_D + m_{K})^2}$, while $(m_{23}^2)_{max}={(E_2^*+E_3^*)^2-(\sqrt{E_2^{*2} - m_{K}^2} - \sqrt{E_3^{*2}-m_{\pi}^2})^2}$ and $(m_{23}^2)_{min}={(E_2^*+E_3^*)^2-(\sqrt{E_2^{*2} - m_{K}^2} + \sqrt{E_3^{*2}-m_{\pi}^2})^2}$. $E_2^* = (m_{12}^2 - m_D^2 + m_{K}^2)/2m_{12}$ and $E_3^* = (m_{B_c}^2 - m_{12}^2 - m_{\pi}^2)/2m_{12}$. We also get nonvanishing contribution in the RPV model from the matrix elements
\begin{eqnarray}
	&&\langle D_s^- K^- \pi^+ | \mathcal O_{R_1} | B_c^- \rangle = -\langle D_s^- K^- \pi^+ | \mathcal O_{R_2} | B_c^- \rangle \nonumber\\*
	&&\quad= \langle \bar K^- \pi^+ | (\bar s (1-\gamma_5) d) | 0 \rangle \langle D_s^-  | (\bar s (1+\gamma_5) b) | B_c^- \rangle \nonumber\\*
	&&\quad= F_0^{BD}(s) F_0^{K\pi}(s) \frac{(m_{B_c}^2-m_{D_s}^2)(m_K^2-m_{\pi}^2)}{(-m_s+m_d)(m_b-m_s)}.
\label{BDKP_R}
\end{eqnarray} 
Finally, THDM  contributes to this decay mode with the matrix element for the $\mathcal O_{T2}$ operator the same as (\ref{BDKP_R}).

\subsubsection{$B_c^- \rightarrow D_s^{*-} K^- \pi^+$ decay}

In SM and MSSM the matrix element of the operator $\mathcal O$ can be factorized as
\begin{eqnarray}
	\langle D_s^{*-} K^- \pi^+| \mathcal O | B_c^- \rangle &=& \langle K^- \pi^+ | (\bar s \gamma^{\mu} (1-\gamma_5) d) | 0 \rangle \nonumber\\*
	&&\times \langle D^*_s  | (\bar s \gamma^{\mu} (1-\gamma_5) b) | B_c^- \rangle.
\end{eqnarray}
The actual expression of the amplitude $\mathcal A$ in terms of the corresponding form factors turns out to be rather long and complicated. This is is even more so for the squared amplitude $\sum_{\mathrm{pol.}}|\mathcal A|^2$, once we have summed over polarizations of the $D_s^{*-}$ meson. Thus, we only give for this mode in SM and MSSM the final numerical values of the branching ratios in the results section. While the THDM does not contribute to this decay mode, the RPV model also gives a non-vanishing contribution from the matrix elements
\begin{eqnarray}
	&&\langle D_s^{*-} K^- \pi^+ | \mathcal O_{R_1} | B_c^- \rangle = \langle D_s^{*-} K^- \pi^+ | \mathcal O_{R_2} | B_c^- \rangle \nonumber\\*
	&&\quad= \langle \bar K^- \pi^+ | (\bar s (1-\gamma_5) d) | 0 \rangle \langle D^*_s  | (\bar s (1+\gamma_5) b) | B_c^- \rangle \nonumber\\*
	&&\quad = -2i \Big[ A_0^{BD}(s) m_{D_s^{*}} - A_1^{BD}(s) (m_{B_c}+m_{D_s^{*}})\nonumber\\*
	&&\quad\quad + A_2^{BD}(s) (m_{B_c}-m_{D_s^{*}}) \Big] p_B \cdot \epsilon_D \nonumber\\*
	&&\quad\quad \times F_0^{K\pi}(s) \frac{(m_K^2-m_{\pi}^2)}{(-m_s+m_d)(m_b+m_s)},
\label{BDsKPR}
\end{eqnarray} 
while summing over polarizations of the $D_s^{*-}$ meson gives $\sum_{\mathrm{pol.}}|p_B \cdot \epsilon_D|^2=\lambda(m_{B_c}^2,s,m_{D_s^{*}}^2)$. Note that expression (\ref{BDsKPR}) involves a slightly different combination of $B_c^-\rightarrow D_s^{*-}$ form factors than for example the analogous expression (\ref{DK_R}). This is because we can assume $2m_{D_s^*} A_0 (m_K^2)\simeq (m_{B_c}+m_{D_s^*}) A_1 (m_K^2) - (m_{B_c}-m_{D_s^*}) A_2(m_K^2)$ at small kaon masses, while the same is not true for all values of $s$ in the kinematic range. 

\subsubsection{$B_c^- \rightarrow D_s^- D_s^{*-} D^+$ decay}

Here we get two contributions to the decay amplitude. 
%(See figure (\ref{DDs_diagram})).
%\begin{figure}
%\begin{center}
%\input{DsDsDDiagrams}
%\end{center}
%\caption{\label{DDs_diagram}Diagrams contributing to the amplitude $B_c^-\rightarrow D_s D_s^{*-} D^+$. The dashed lines represent vector, pseudoscalar and scalar $K$ meson resonance exchange.}
%\end{figure} 
The matrix element for the operator $\mathcal O$ becomes 
\begin{eqnarray}
	&&\langle D_s^- D_s^{*-} D^+ | \mathcal O | B_c^- \rangle \nonumber\\*
	&&\quad= \langle D_s^- | \bar s \gamma^{\mu}(1-\gamma^5) b | B_c^- \rangle \langle D_s^{*-} D^+ | \bar s \gamma_{\mu} (1-\gamma_5) d | 0 \rangle \nonumber\\*
	&&\quad\quad + \langle D^{*-}_s | \bar s \gamma^{\mu}(1-\gamma^5) b | B_c^-\rangle \langle D_s^- D^+ | \bar s \gamma_{\mu} (1-\gamma_5) d | 0 \rangle \nonumber \\*
	&&\quad= (m_{B_c}^2-m_{D_s^*}^2)F_-^{A D_s^{*-}D}(q^2) F_0^{B D}(q^2) \varepsilon_{D_s^{*-}} \cdot q \nonumber\\*
	&&\quad\quad+ (m_{D_s}^2-m_{D}^2)F_-^{ABD^*}(q^2) F_0^{D_sD}(q^2)  \varepsilon_{D_s^{*-}} \cdot q  \nonumber\\*
\end{eqnarray} 
The second term turns out to be relatively suppressed due to small values of $f_{K^*(1430)}$ and $g_{\pi}$, in the $F_0^{D_sD}$ form factor while averaging over polarizations in the first term gives $\sum | \epsilon^*_{D_s} \cdot q |^2 = \lambda(q^2,m_D^2,m_{D_s}^2)$. Similarly in the RPV model
we only keep the first of the two possible amplitudes $\langle D_s^- D_s^{*-} D^+ | \mathcal O_{R_1} | B_c^- \rangle = \langle D_s^- D_s^{*-} D^+ | \mathcal O_{R_2} | B_c^- \rangle \simeq \langle D_s^- | \bar s (1-\gamma^5) b | B_c^- \rangle \times \langle D_s^{*-} D^+  | \bar s  (1+\gamma_5) d | 0 \rangle$, so that 
\begin{eqnarray}
	&&\langle D_s^- D_s^{*-} D^+ | \mathcal O_{R_1} | B_c^- \rangle = F_0^{BD}(q^2) F_-^{AD^*_sD}(q^2) \nonumber\\*
	&&\quad\times\frac{(m_{B_c}^2 - m_{D_s}^2) q^2}{(m_s+m_d)(m_b-m_s)} \epsilon^*_{D_s} \cdot q.
\end{eqnarray}

\subsubsection{$B_c^- \rightarrow D_s^- D_s^- D^{*+}$ decay}

The matrix element of the $\mathcal O$ operator becomes in this case 
\begin{eqnarray}
	&&\langle D_s^- D_s^- D^{*+} | \mathcal O | B_c^- \rangle = \langle D_s^-(p_{D_1}) | \bar s \gamma^{\mu}(1-\gamma^5) b | B_c^-(p_B) \rangle \nonumber\\*
	&&\quad\times \langle D_s^-(p_{D_2}) D^{*+}(p_{D_3}) | \bar s \gamma_{\mu} (1-\gamma_5) d | 0 \rangle \nonumber\\*
	&&\quad+ (p_{D_1} \leftrightarrow p_{D_2}),
\end{eqnarray}
so that $\sum_{\mathrm{pol.}}|\mathcal M|^2$ $=$ $| C_{SM}/4 |^2$ 
$\times$ $f(t) ( 2 f(s)(m_{D^*}^2$ $-$ $m_{D_s}^2) $ $+$ $f(t)$ $(t-u) )$, where $f(x) = (m_{B_c}^2 $ $-$ $ m_{D_s}^2)$ $\times$ $ F_-^{A D_sD^*}(x) F_0^{BD}(x)$. In the RPV model we get for this mode equal matrix elements of operators $\mathcal O_{R_1}$ and $\mathcal O_{R_2}$
\begin{eqnarray}
	&&\langle D_s^- D_s^- D^{*+} | \mathcal O_{R_1} | B_c^- \rangle = \langle D_s^- D_s^- D^{*+} | \mathcal O_{R_2} | B_c^- \rangle \nonumber\\*
	&&\quad= \langle D_s^-(p_{D_1}) | \bar s (1-\gamma^5) b | B_c^-(p_B) \rangle \nonumber\\*
	&&\quad\quad \times \langle D_s^-(p_{D_2}) D^{*+}(p_{D_3}) | \bar s  (1+\gamma_5) d | 0 \rangle \nonumber\\*
	&& \quad\quad+ (p_{D_1} \leftrightarrow p_{D_2}),
\end{eqnarray} 
so that $\sum_{\mathrm{pol.}} | \mathcal M |^2 =( f_{QCD} / m_{\widetilde \nu}^2 )^2 \times ( \sum_n | \lambda'_{n32} \lambda_{n21}'^* |^2 + | \lambda'_{n21} \lambda_{n32}'^* |^2 ) \times g(t)  ( 2 g(s)( m_{D^*}^2 - m_{D_s}^2 ) + g(t) ( t - u ) )$, where $g(x) = ( m_{B_c}^2 - m_{D_s}^2 ) \times F_-^{A D_sD^*}(x) F_0^{BD}(x) \times x / ( m_s + m_d )( m_b - m_s )$.

\subsubsection{$B_c^- \rightarrow D_s^- D_s^- D^{+}$ decay}

The matrix element of the operator $\mathcal O$ is calculated to be 
\begin{widetext}
\begin{eqnarray}
	\langle D_s^- D_s^- D^{+} | \mathcal O | B_c^- \rangle &=& \langle D_s^-(p_{D_1}) | \bar s \gamma^{\mu}(1-\gamma^5) b | B_c^-(p_B) \rangle \times \langle D_s^-(p_{D_2}) D^{+}(p_{D_3}) | \bar s \gamma_{\mu} (1-\gamma_5) d | 0 \rangle \nonumber\\
&& + (p_{D_1} \leftrightarrow p_{D_2}) \nonumber\\
&=& f_+^{BD}(s) f_+^{D_sD}(s) (t-u) + f_+^{BD}(s) f_-^{D_sD}(s) ( m_{D_s}^2 - m_{B_c}^2 )\nonumber\\
&& + f_-^{BD}(s) f_+^{D_sD}(s) (m_{D}^2 - m_{D_s}^2) + f_-^{BD}(s) f_-^{D_sD}(s) (-s)\nonumber\\
&& + (s \leftrightarrow t)
\end{eqnarray}
\end{widetext}
RPV model also allows for this mode 
\begin{eqnarray}
	&&\langle D_s^- D_s^- D^{+} | \mathcal O_{R_1} | B_c^- \rangle = \langle D_s^- D_s^- D^{+} | \mathcal O_{R_2} | B_c^- \rangle \nonumber\\*
	&&\quad= \langle D_s^-(p_{D_1}) | \bar s (1-\gamma^5) b | B_c^-(p_B) \rangle \nonumber\\*
	&&\quad\quad \times \langle D_s^-(p_{D_2}) D^{+}(p_{D_3})(q) | \bar s  (1+\gamma_5) d | 0 \rangle \nonumber\\*
	&&\quad\quad+ (p_{D_1} \leftrightarrow p_{D_2}) \nonumber\\*
	&&\quad= F_0^{BD}(s) (m_{B_c}^2-m_{D_s}^2) F_0^{D_sD}(s) \nonumber\\* 
	&&\quad\quad\times\frac{(m_{D_s}^2-m_{D}^2)}{(m_s-m_d)(m_b-m_s)} + (s \leftrightarrow t).
\label{DDDs_R}
\end{eqnarray} 
Finally, the THDM contributes to the amplitude of this decay through the operator $\mathcal O_{T_1}$ giving the same result as (\ref{DDDs_R}).

\subsubsection{$B_c^- \rightarrow \bar K^0 D^0 K^{-}$ decay}

Also this transition has nonzero amplitudes in all models considered. We analyzed this decay amplitude with the assumption of $D_s^{*-}$ vector meson resonance exchange and used HQET for the evaluation of the $D_s^{*-} D^0 K$ vertex. The matrix element of the operator $\mathcal O$ turns out to be
\begin{eqnarray}
	&&\langle \bar K^0 D^0 K^{-} | \mathcal O | B_c^- \rangle \nonumber\\*
	&&\quad= \langle D^{*-}_s(q,\epsilon^{\mu}) | \bar s \gamma^{\mu}(1-\gamma^5) b | B_c^-(p_B) \rangle \nonumber\\*
	&&\quad\quad\times\langle \bar K^0 (p_{K1}) | \bar s \gamma_{\mu} (1-\gamma_5) d | 0 \rangle \nonumber\\*
	&&\quad\quad\times i D_{\mu\nu}^{D^*_s}(q) \times\langle \bar D^0(p_{D}) K^{-}(p_{K2}) |\mathcal L_{S}| D^{*-}_s(q,\epsilon^{\nu})\rangle,\nonumber\\*
\end{eqnarray}
where $D_{\mu\nu}^{D_s^{*-}}(q)=(-g_{\mu\nu} + q_{\mu}q_{\nu}/m_{D_s^*}^2)/(q^2 - m_{D_s^*}^2 + i \Gamma(D_s^{*}) m_{D_s^*})$ is the $D_s^{*-}$ meson propagator while  $\langle \bar D^0(p_{D}) K^{-}(p_{K2}) |\mathcal L_{S} |  D^{*-}_s(q) \rangle = (2g_{A}/f_{K})\times\sqrt{m_{D_s^*} m_{D}} \times \epsilon^*_D\cdot p_{K2}$ is the strong transition matrix element calculated in HQET. The amplitude thus becomes
\begin{eqnarray}
	&&\mathcal M(B_c^- \rightarrow \bar K^0 D^0 K^{-}) =\frac{1}{4} C_{SM}  2 i f_K A^{BD}_0(m_K^2) \nonumber\\*
	&&\quad\times \frac{(s + m_{D_s^*}^2) (u - 2 m_{K}^2)}{2 m_{D_s^*}^2 
(s + i \Gamma(D_s^{*}) m_{D_s^*} - m_{D_s^*}^2)} i 2 \frac{g_A}{f_{K}} \sqrt{m_{D_s^*} m_{D}}.\nonumber\\*
\end{eqnarray}
Similarly in the RPV model we get
\begin{eqnarray}
	&&\langle \bar K^0 D^0 K^{-} | \mathcal O_{R_1} | B_c^- \rangle = \langle \bar K^0 D^0 K^{-} | \mathcal O_{R_2} | B_c^- \rangle \nonumber\\*
	&&\quad= \langle D^{*-}_s(q,\epsilon^{\mu}) | \bar s (1-\gamma^5) b | B_c^-(p_B) \rangle \nonumber\\*
	&&\quad\quad\times \langle \bar K^0(p_{K1}) | \bar s  (1+\gamma_5) d | 0 \rangle \nonumber\\*
	&&\quad\quad\times i D_{\mu\nu}^{D^*_s}(q) \times \langle \bar D^0(p_{D}) K^{-}(p_{K2}) |\mathcal L_{S}| D^*_s(q,\epsilon^{\nu}) \rangle\nonumber\\*
	&&\quad= \frac{2 i f_K m_K^2 A^{BD}_0(m_K^2)}{(m_b+m_s)(m_d+m_s)} \nonumber\\*
	&&\quad\quad\times \frac{(s + m_{D_s^*}^2) (u - 2 m_{K}^2 )}{2 m_{D_s^*}^2 
(s + i \Gamma(D_s^{*}) m_{D_s^*} - m_{D_s^*}^2)} i 2 \frac{g_A}{f_{\pi}} \sqrt{m_{D_s^*} m_{D}}.\nonumber\\*
\label{DKK_R}
\end{eqnarray}
Finally in the THDM only the operator $O_{T_1}$ has non-vanishing amplitude and gives the same result as (\ref{DKK_R}).

\subsubsection{$B_c^- \rightarrow \bar K^0 D^{*0} K^-$ decay}

This decay model was analyzed by presuming $D_s^-$ pseudoscalar meson resonance exchange. The matrix element of the operator $\mathcal O$ was found to be
\begin{eqnarray}
	&&\langle \bar K^0 D^{*0} K^{-} | \mathcal O | B_c^- \rangle \nonumber\\*
	&&\quad= \langle D_s^-(q) | \bar s \gamma^{\mu}(1-\gamma^5) b | B_c^-(p_B) \rangle \nonumber\\*
	&&\quad\quad\times\langle \bar K^0 (p_{K1}) | \bar s \gamma_{\mu} (1-\gamma_5) d | 0 \rangle \nonumber\\*
	&&\quad\quad\times i \Delta^{D_s}(q) \times\langle \bar D^{*0}(p_{D}) K^{-}(p_{K2}) |\mathcal L_{S}| D_s(q)\rangle,\nonumber\\*
\end{eqnarray}
where we denoted the pseudoscalar $D_s^-$ meson propagator with $\Delta_F^{D_s}(q)=1/(q^2-m_{D_s}^2)$. Evaluating the strong interaction vertex with HQET we get for the amplitude
\begin{eqnarray}
	&&\mathcal M(B_c^- \rightarrow \bar K^0 D^{*0} K^{-}) =\frac{1}{4} C_{SM}  f_K F_0^{BD}(m_K^2)\frac{m_{B_c}^2-m_{D_s}^2}{q^2-m_{D_s}^2} \nonumber\\*
	&&\quad\times i 2 \frac{g_A}{f_{K}} \sqrt{m_{D_s} m_{D^*}} \epsilon^*_D \cdot p_{K2}.\nonumber\\*
\end{eqnarray}
RPV model also contributes to this decay mode, giving for the $\mathcal O_{R1}$ and $\mathcal O_{R2}$ matrix elements
\begin{eqnarray}
	&&\langle \bar K^0 D^{*0} K^{-} | \mathcal O_{R_1} | B_c^- \rangle = -\langle \bar K^0 D^{*0} K^{-} | \mathcal O_{R_2} | B_c^- \rangle \nonumber\\*
	&&\quad= \langle D_s^-(q) | \bar s (1-\gamma^5) b | B_c^-(p_B) \rangle \nonumber\\*
	&&\quad\quad\times \langle \bar K^0(p_{K1}) | \bar s  (1+\gamma_5) d | 0 \rangle \nonumber\\*
	&&\quad\quad\times i \Delta^{D_s}(q) \times \langle \bar D^{*0}(p_{D}) K^{-}(p_{K2}) |\mathcal L_{S}| D^-_s(q) \rangle\nonumber\\*
	&&\quad= \frac{f_K m_K^2 F_0^{BD}(m_K^2)}{(m_b-m_s)(m_d+m_s)} \nonumber\\*
	&&\quad\quad\times \frac{m_{B_c}^2-m_{D_s}^2}{q^2-m_{D_s}^2} i 2 \frac{g_A}{f_{K}} \sqrt{m_{D_s} m_{D^*}} \epsilon^*_D \cdot p_{K2}.\nonumber\\*
\end{eqnarray}

\section{Results and Discussion}

The usefulness of $\Delta S = 2$ decays of $B$ mesons in the search for new physics has been discussed in several publications~\cite{Huitu:1998vn,Huitu:1998pa,Grossman:1999av,Fajfer:2001ht,Chun:2003rg,Wu:2003kp,Fajfer:2000ax,Fajfer:2000ny}. From the models considered so far it appears that these decays are particularly relevant in the search for supersymmetry, with and without $\mathcal R$-parity violation. In the present paper we have undertaken a similar investigation with regard to the $\Delta S = 2$ decays of the $B_c$ meson.
\par
The calculation of matrix elements has been performed in the factorization approximation. In the $B$-meson decays this works in a good number of cases, while in other cases a more sophisticated approach is needed (for a recent review see~\cite{Wei:2003tv}). In this first calculation we considered that factorization approximation is sufficient for obtaining the correct features of the decays of the various channels considered. An exception is the case in which matrix elements vanish as a result of factorization, which in a better approximation can be improved.
\par
The results obtained in the MSSM framework depend on the values of the $\delta_{ij}^{d}$ parameters of the mass-insertion approximation which we use. The constraints on these parameters have been improved in recent years~\cite{Ciuchini:2002uv,Ciuchini:2003rg} vs. the values which were used in the original calculation~\cite{Huitu:1998vn} of the $b \rightarrow s s \bar d$.
%as a function of $x=m_{\tilde g}^{2}/m_{\tilde d}$ in the MSSM without $\mathcal R$-parity violation, and we present the result in Fig. \ref{xgraph}. 
%\begin{figure}
%\includegraphics[height=300pt,angle=-90]{brplot.eps}
%\input{brplot}
%\caption{\label{xgraph} The inclusive branching ratio of $b \rightarrow s s /bar d$ as a function of $x=m_{\tilde g}^2/m_{\tilde d}^2$ in MSSM without $\mathcal R$-parity violation.}
%\end{figure} 
%This figure is therefore an update of Fig. 1 of reference~\cite{Huitu:1998vn}. 
For the $\mathcal R$-parity violating MSSM we use the new limits on the $\lambda_{ij}'$ couplings of the superpotential, which we obtain by the use of the new limits on the $B^{-} \rightarrow K^- K^- \pi^+$ decay from BELLE~\cite{Garmash:2003er} and the explicit calculation of this decay of reference~\cite{Fajfer:2001te}.
\par
In Table \ref{results_table} we summarize the results of our calculation for the various possible decay modes considered. 
%\begingroup
\begin{table*}
\begin{center}
\begin{ruledtabular}
\begin{tabular}{lccccc}
Decay & SM & MSSM $(x=1)$ & MSSM $(x=8)$ & RPVM & THDM \\
\hline
$B_c^- \rightarrow D_s^{*-} \bar K_0^*$ & $8 \times 10^{-14}$ & $5\times 10^{-13}$ & $2\times 10^{-9}$ & - & - \\
$B_c^- \rightarrow D_s^{*-} \bar K_0$ & $5 \times 10^{-14}$ & $3\times 10^{-13}$ & $1\times 10^{-9}$ & $1 \times 10^{-5}$ & $10^{-10}$ \\
$B_c^- \rightarrow D_s^- \bar K_0^*$ & $6 \times 10^{-15}$ & $4 \times 10^{-14}$ & $2\times 10^{-10}$ & - & - \\
$B_c^- \rightarrow D_s^- \bar K_0$ & $4 \times 10^{-15}$ & $2 \times 10^{-14}$ & $9 \times 10^{-11}$ & $1 \times 10^{-6}$ & - \\
$B_c^- \rightarrow D_s^- K^- \pi^+$ & $2 \times 10^{-14}$ & $1\times 10^{-13}$ & $4\times 10^{-10}$ & $6 \times 10^{-5}$ & $10^{-10}$ \\
$B_c^- \rightarrow D_s^{*-} K^- \pi^+$ & $6 \times 10^{-14}$ & $4\times 10^{-13}$ & $1\times 10^{-9}$ & $9 \times 10^{-5}$ & - \\
$B_c^- \rightarrow D_s^- D_s^{*-} D^+$ & $3 \times 10^{-17}$ & $2\times 10^{-16}$ & $7\times 10^{-13}$ & $5 \times 10^{-5}$ & - \\
$B_c^- \rightarrow D_s^- D_s^- D^{*+}$ & $1 \times 10^{-17}$ & $7\times 10^{-17}$ & $3\times 10^{-13}$ & $2 \times 10^{-5}$ & - \\
$B_c^- \rightarrow D_s^- D_s^- D^{+}$ & $9 \times 10^{-20}$ & $6\times 10^{-19}$ & $2\times 10^{-15}$ & $2 \times 10^{-7}$ & $10^{-12}$  \\
$B_c^- \rightarrow D_0 \bar K_0 K^{-}$ & $2 \times 10^{-14}$ & $1\times 10^{-13}$ & $6\times 10^{-10}$ & $6 \times 10^{-6}$ & $10^{-11}$ \\
$B_c^- \rightarrow D_0^* \bar K_0 K^{-}$ & $2 \times 10^{-14}$ & $1\times 10^{-13}$ & $4\times 10^{-10}$ & $5 \times 10^{-6}$ & - \\
\end{tabular} 
\end{ruledtabular}
\caption{The predicted branching ratios for the $\Delta S= 2$ two- and three-body decays of $B_c^-$ calculated using the factorization approach within Standard model (the first column), Minimal supersymmetric standard model for $x=1$ and $x=8$ as explained in the text (the second and third column), Minimal supersymmetric standard model extended by ${\cal R}$- parity breaking (the fourth column), and two Higgs doublet model (the fifth column). In columns (2)-(5) the calculated upper limits are given as allowed by presently known limits from other processes.\label{results_table}}
\end{center} 
\end{table*} 
%\endgroup
As expected, the SM results for all these $\Delta S = 2$ channels are extremely small. For the models beyond SM, the values presented are the upper limits presently allowed by the constraints available from other processes. For MSSM we present the results for the two extreme values of $x=1$ and $8$. We note that MSSM with $\mathcal R$-parity violation is rather poorly restricted at present for some of the decays, in particular the decays  $B_c^- \rightarrow D_s^{*-} \bar K^0$, $B_c^- \rightarrow D_s^{*-} K^- \pi^+$, $B_c^- \rightarrow D_s^- K^- \pi^+$, $B_c^- \rightarrow D_s^- D_s^{*-} D^+$, $B_c^- \rightarrow D_s^- D_s^{-} D^{*+}$, which are in the $10^{-5}-10^{-4}$ range. All these decays should be looked for, when sizable samples of $B_c$'s will be available.

\begin{acknowledgments}
We would like to thank B. Golob for very useful discussions on experimental aspects of this investigation. S.F. and J.K. are supported in part by the Ministry of Education, Science and Sport of the Republic of Slovenia.
\end{acknowledgments}

\bibliography{text}

\end{document}